\documentclass[preprint,5p,twocolumn,numbers]{elsarticle}

\usepackage{amssymb}
\usepackage{amsmath}
\usepackage[colorlinks = true,
  linkcolor = blue,
  citecolor = blue,
  urlcolor = blue]{hyperref}
\usepackage{balance}
\usepackage{graphicx}
\usepackage{booktabs}
\usepackage{multirow}
\usepackage{tabularx}
\usepackage{enumitem}
\usepackage{threeparttable}
\usepackage{url}

\begin{document}

\title{A Bird's-Eye View on Security Considerations in RFCs}
\author[sdu]{Jukka Ruohonen\corref{cor}}
\ead{juk@mmmi.sdu.dk}
\author[sdu]{Qusai Ramadan}
\cortext[cor]{Corresponding author.}
\address[sdu]{University of Southern Denmark, S\o{}nderborg, Denmark}

\begin{abstract}
Request for comments (RFCs) are Internet standards, memorandums, and related
technical documents about core Internet protocols made via and released by the
Internet Engineering Task Force (IETF). In the early 1990s each RFC was required
to have a section for security considerations. The present work examines these
sections. According to the empirical results, (1)~over 90\% of the RFCs sampled
have discussed security explicitly in these sections, (2)~although mandatory
security requirements have only seldom---if ever---been imposed. Furthermore,
(3)~the RFC-to-RFC reference network specific to the security consideration
sections is sparse, although a few RFCs and their security consideration
sections are heavily referenced. In addition, (4) the volume of references
peaked during a period from circa mid-1990s to mid-2010s. Regarding the topics
discussed in the sections, (5) these do not represent general security issues,
such as spoofing or eavesdropping; rather, the topics mostly reflect distinct
security issues specific to distinct protocols. With the exceptions of network
security in general, security specifications, and routing, (6) also the
longitudinal evolution of the topics is protocol-specific. As the subject matter
has not been previously examined, these empirical results fill a gap in the
standardization literature.
\end{abstract}


\maketitle

\section{Introduction}

The Internet's early design philosophy emphasized multiplexing, reliability,
heterogeneity of networks and hosts, distributed management, cost-effectiveness,
scalability, and accountability~\cite{Clark88}. Security was not an explicit
concern. It was only later on, in the 1990s, when security started to appear as
a design principle in conjunction with the end-to-end architectural design
principle~\cite{RFC1958}. It was also in the early 1990s when all RFCs were
mandated to have explicit sections for discussing any and all security
considerations that a new protocol, memorandum, or other technical document
might carry~\cite{RFC1543}. The present work examines these explicit security
considerations. What do they say? What security domains and concepts are
pronounced? Which RFCs are highly referenced in terms of security? Are security
requirements imposed? These motivating questions can be formalized into the following four research questions (RQs):
\begin{itemize}
\item{RQ.1: How many RFCs have explicitly acknowledged security considerations,
  and are IETF's standardized keywords for requirements used in these?}
\item{RQ.2: How many RFCs are central references in terms of explicitly
  acknowledged security considerations?}
\item{RQ.3: When were the central RFC references published; how has the
  referencing evolved over time?}
\item{RQ.4: What are the main topics discussed in the explicitly acknowledged
  security considerations of RFCs, and how have the topics evolved over time?}
\end{itemize}

 As soon discussed in the opening Section~\ref{sec: background}, these and
 related questions have neither been asked nor answered previously. Though,
 before continuing, a clarifying remark is required about the questions: even
 though the security considerations sections can be viewed as a proxy for how
 security considerations are documented within the IETF, the results should not
 be interpreted as characterizing the complete security reasoning or discussions
 that took place during protocol design and standardization work. Rather, they
 characterize how security considerations have been documented in RFCs. Then,
 after presenting the dataset examined in Section~\ref{sec: data}, the methods
 are elaborated in Section~\ref{sec: methods}. The empirical results are
 subsequently presented in Section~\ref{sec: results}. The conclusion, including
 the answers to the research questions, is presented in the last
 Section~\ref{sec: conclusion}. It also provides a brief discussion about
 limitations, practical implications, and further research directions.

\section{Background and Related Work}\label{sec: background}

The IETF and its history have been extensively studied, including with respect
to RFCs and the work practices for standardization around them~\cite{Coleman13,
  Welzl21}. Of the existing work, particularly noteworthy is the research on the
adoption of protocols and the adoption challenges sometimes
encountered~\cite{Leva13, McQuistin21}. Although security surely is a challenge
for the Internet and its protocols, it may paradoxically not be a challenge for
the adoption of new protocols.

As an informational RFC from the late 2000s noted, protocols ``with security
flaws may still become wildly successful provided that they are extensible
enough to allow the flaws to be addressed in subsequent
revisions''~\cite{RFC5218}. The RFC further contemplated that security
vulnerabilities ``do not seem to limit initial success, since vulnerabilities
often become interesting to attackers only after the protocol becomes widely
deployed enough to become a useful target''~\cite{RFC5218}. Rather analogous
reasoning is behind a so-called popularity hypothesis in software
security~\cite{Ruohonen26REP}. In fact, implicitly, one finds analogous
reasoning already from the very first RFC 1 released in 1969; ``there will be a
period of very light usage until the community of users experiments with the
network and begins to depend upon~it''~\cite{RFC1}. When it comes to security,
the later ``experimentation'' since the 1990s was not necessarily what was
envisioned in the late \text{1960s---even} though the point about
experimentation in itself certainly stood the test of time.

Furthermore, somewhat similar sentiments have been expressed by academic
researchers having been involved at the IETF's standardization; the Internet's
``protocols are a convoluted mess, patched together to solve numerous short-term
issues'', but this state of affairs ``is expected and
normal''~\cite[p.~6]{Weltzl23}. However, when it comes to security-specific
techniques, practices, and standards, adoption challenges have frequently been
discussed in the literature~\cite{Ruohonen25JISA, Ruohonen26SANER}. The IETF's
protocols for the Internet are not an exception. The domain name system is a
good example; despite various new security improvements, adoption of them has
been slow and various sophisticated attacks continue~\cite{Schmid21,
  Yajima21}. Although patching is possible and even a normal practice with
Internet protocols, a strong argument can thus be raised that with security, in
particular, diligence early on may well pay off. Fixing things later on can be
difficult and costly.

Finally, it is worth pinpointing a few works that are related through
methodology. In particular, the citations in RFCs and the keywords used in them
have been examined also previously~\cite{Lin23, McQuistin21}, including with
network analysis~\cite{Gencer12} used also in the present work. However: to the
best of the authors' knowledge based on thorough but still non-systematic
literature searches, none of the methodologically related previous works have
explicitly concentrated on the security considerations expressed in RFCs. There
are some implicitly related works that have analyzed these in distinct domains
such as Internet-of-things~\cite{Morabito20}. Also the IETF's work on privacy
has been surveyed~\cite{Dikshit23}.  Again, however, neither have these
contextually related works considered the security considerations as a
whole. Therefore, the present work fills a gap in existing research.

\section{Data}\label{sec: data}

\subsection{Retrieval}

The data was collected on 15 March 2026 from the IETF's online
archive.\footnote{~\url{https://www.rfc-editor.org/rfc/}} In practice, the
collection involved a loop $i = 1, \ldots, m$ with which both obsoleted RFCs and
RFCs in-force were retrieved sequentially. For instance: when \text{$i = 2000$},
the corresponding RFC was retrieved from the archive as a \texttt{rfc2000.html}
file. To ease parsing, the files were retrieved in the hypertext markup language
format. Given the data collection date, the maximum was set to $m = 10,000$
because the exact amount of RFCs was not explicitly known at the time of data
collection. Although the IETF indeed uses sequential numbers for RFCs, a few
older ones were not available from the archive; in total, $290$ such cases were
recorded during the retrieval. This amount includes also a few RFCs that were
not yet published at the time of the data collection; RFC~9939 is the latest
request for comments observed.

\subsection{Parsing}\label{subsec: parsing}

The analysis is based on the specific security considerations section that was
first introduced in RFC 1543. According to this RFC from 1993, all ``RFCs must
contain a section near the end of the document that discusses the security
considerations of the protocol or procedures that are the main topic of the
RFC''~\cite{RFC1543}. In total, $1,1270$ RFCs did not have such a section, which
is understandable due to the looping from the very first RFC that was published
as early as 1969. When the downloading errors are further excluded, the analysis
operates with $n = 8,440$ RFCs having explicit sections for security
considerations. Thus, given the numbers mentioned, $m = 290 + 1,270 + n$.

\begin{table}[th!b]
\centering
\caption{Strings for the Absence of Explicit Security Considerations}
\label{tab: string matching}
\begin{scriptsize}
\begin{tabularx}{\linewidth}{X}
\toprule
``this document does not impact the security'' \\
``security is not explicitly discussed in this'' \\
``Security is not discussed in this'' \\
``Security is not addressed in this'' \\
``Security is not directly addressed by this'' \\
``Security is not discussed directly in this'' \\
``Security isses are not discussed in this'' \\
``Security issues are not discussed in this'' \\
``Security issues are not addressed in this'' \\
``Security issues are not considered in this'' \\
``Security concerns are not addressed in this'' \\
``Security considerations are not addressed here'' \\
``Security considerations are not addressed in this'' \\
``Security considerations are not discussed in this'' \\
``Security issues are not directly discussed in this'' \\
``Security issues are not discussed in detail in this'' \\
``Security issues are not explicitly discussed in this'' \\
``Security issues are not specifically addressed in this'' \\
``Security issues are not substantially discussed in this'' \\
``This RFC does not discuss security issues'' \\
``This RFC does not address issues of security'' \\
``This document does not affect the security issues'' \\
``This document does not alter the security properties'' \\
``This update does not change the security considerations'' \\
``This memo does not directly address security issues'' \\
``This document does not discuss security considerations'' \\
``This document does not introduce additional security requirements'' \\
``This document does not raise any additional security issue'' \\
``Security and Liability issues are not discussed in this'' \\
\bottomrule
\end{tabularx}
\end{scriptsize}
\end{table}

However, having a standardized section for security considerations does not mean
that security is always addressed in the RFCs. This point is also why the RQs
emphasize the adjective \textit{explicit} in conjunction with the security
considerations. Regarding RQ.1, two simple parsing rules are used: a RFC's
section for security considerations is taken to lack explicit security
consideration in case (a) the section's length is less than thirty characters or
(b) any of the strings in Table~\ref{tab: string matching} appear anywhere in
the section.

The latter part of RQ.1 needs a brief comment too. Although RFC 8174 clarified
that the IETF's standardized keywords for requirements, including with respect
to cases such as MAY, MUST, and MUST NOT, should only be considered when spelled
in capital letters~\cite{RFC8174}, lower-cased matches are counted for the
eleven keywords specified in the RFC mentioned. The reason is simple: none of
the RFCs sampled specified even a single upper-cased keyword in their sections
for security~considerations.

It should be remarked that only the security sections are parsed. If used, the
parsing includes subsections present in the main security considerations
section; see \cite{RFC4187} for an example about such cases. However, those RFCs
are not fully assessed that have placed their security discussions into
appendices that are then referenced in the brief security sections; see
\cite{RFC8446} for an example. Nor are references followed. In other words, a
RFC whose section for security considerations references other RFCs is parsed
only in terms of what is explicitly written in the section. These choices are
sensible because attempting to cover complex within-RFC and between-RFC
referencing would presumably increase a probability of parsing errors.

\section{Methods}\label{sec: methods}

\subsection{Network Analysis}\label{subsec: network analysis}

Network (graph) analysis is used for answering to RQ.2. Like with citation
networks in science, a weighted and directed network is used; vertices refer to
RFCs and edges to references between them. More specifically, a vertex in the
network means that a given RFC either referenced other RFCs in its section for
security considerations or the RFC was referenced by other RFCs in their own
sections for security considerations. Thus, the network is specific to RFCs;
other references, including references to other standards, are excluded. Also
within-RFC references are excluded; hence, there are no self-loops in the
weighted and directed network. Analogously to citations in science, edge weights
denote reference counts. It should also be remarked that the RFCs without
explicit security considerations (cf.~RQ.1) are included in the network
insofar as they referenced other RFCs or were referenced by other~RFCs.

The wording in RQ.2 and RQ.3 about central references is approached with three
conventional network centrality measures: weighted in-degree centrality,
betweenness centrality, and closeness centrality. All are well-known centrality
measures needing no particular elaboration; details and definitions are
available from the literature~\cite{Diallo16, Landherr10}. The only exception is
the weighted in-degree centrality for which a concept of strength has sometimes
been used as a synonym~\cite{Barrat04}. An example clarifies things: if a given
RFC only received references from a single another RFC but this referencing RFC
made ten references to the receiving RFC, the weighted in-degree (strength) of
the referenced RFC is ten (and not one, as would be the case with unweighted
in-degree). To also recall: all references between the RFCs observed refer to
those made within their standardized security consideration sections.

In general, only the weighted in-degree centrality conveys a straightforward
interpretation: for a given RFC, it is the count of references made to the RFC
in the security consideration sections of other RFCs. Thus, it can be
interpreted in a similar vein than citations of scientific
publications. Although some interpretation guidance is available also for the
other two centrality measures in scientific citation networks~\cite{Abbasi11,
  Diallo16}, it is difficult to transfer them to the RFC context. Thus, the
answer to RQ.2 is sought by focusing primarily on the weighted in-degree
centrality and secondarily on those other two centrality measures correlated
with the weighted in-degree centrality values. Because many networks are
scale-free, meaning that their degree distributions tend to follow a power law,
Spearman's rank correlation coefficient has been recommend over the Pearson's
product-moment correlation coefficient~\cite{Shao18}. It is thus adopted for
evaluating the correlations between the three centrality measures.

To help answering to RQ.2 and RQ.3, a cutoff point is required. Again due to the
scale-free nature of many networks, a scree test~\cite{Cattell66}, or a scree
plot, which is commonly used dimension reduction computations, has been adopted
for determining suitable cutoff points for centrality values~\cite{Himelboim19,
  Himelboim17}. Although somewhat subjective, a scree plot is thus used for
sharpening the answers to RQ.2~and~RQ.3.

\subsection{Topic Modeling}

The topic modeling is conducted with BERTopic~\cite{BERTopica}. It uses
transformers and class-based term frequency-inverse document frequency
(c-TF-IDF) weights. When compared to the conventional latent Dirichlet
allocation method for topic modeling~\cite{Blei12}, BERTopic is beneficial due
to its leverage of sentence embeddings that often yield semantically richer and
more coherent topic groupings~\cite{Alsulami26}. Although RFCs are technical
documents, they exhibit a considerable lexical diversity through the use of
specialized terminology, synonyms, abbreviations, and related concepts. This
lexical diversity makes BERTopic well-suited for clustering a heterogeneous
corpus such as the security considerations sections of RFCs.  With respect to
the second part of RQ.4, the implementation used~\cite{BERTopica, BERTopicb}
also allows effortless longitudinal evaluation.

With respect to computational details, the number of topics was restricted to $t
= 20$. As is common in topic modeling~\cite{Rijcken22}, the choice was made
primarily based on human interpretability of the topics according to the tokens
in them and their probabilities used as weights. In addition, perplexity values
were secondarily evaluated; lower values are better~\cite{Zhao15}. Each topic
was specified to have at least five tokens. For longitudinal modeling,
evolutionary tuning was used. These configurations notwithstanding, the default
parameters of the implementation were used.

Regarding pre-processing, (1)~the sections for security considerations were
tokenized according to white space and punctuation characters. Then, (2) after
lower-casing the tokens, (3) only alphabetical tokens recognized as nouns were
included. In addition, (4)~tokens with lengths less than three characters or
over twenty characters were excluded alongside stopwords. Finally, (5) a few
stopwords were excluded.\footnote{~In addition to the stopwords present in the
library used for the pre-processing~\cite{NLTK25}, the following custom
stopwords were excluded: \textit{amount}, \textit{document}, \textit{dot},
\textit{end}, \textit{far}, \textit{get}, \textit{going}, \textit{like},
\textit{make}, \textit{med}, \textit{name}, \textit{none}, \textit{note},
\textit{paper}, \textit{part}, \textit{pose}, \textit{publication},
\textit{rfc}, \textit{regard}, \textit{section}, \textit{set}, \textit{single},
\textit{sub}, \textit{thus}, \textit{take}, \textit{us}, \textit{usage},
\textit{use}, \textit{user}, \textit{using}, and \textit{well}. These custom
stopwords were excluded because they made the topic groupings more incoherent
and harder to interpret. In addition, it should be remarked that the library
categorizes also tokens it does not recognize grammatically as
nouns; abbreviations are a good example.} The tokens were finally (6) lemmatized
into their English dictionary forms whenever possible.

A small qualitative cross-check was done by the authors about the interpretation
and labeling of the topics extracted. Namely: after the first author had labeled
and interpreted the topics, including writing of Subsection~\ref{subsec: topic
  modeling results}, the second author was asked to evaluate the labels and
interpretations. All disagreements were recorder. Both authors were further
asked to independently rank each topic with respect to its coherence and
interpretability with a scale from zero (extremely incoherent and impossible to
interpret) to ten (highly coherent and easy to interpret). Regarding the initial
labeling and interpretation made by the first author, when in doubt, the
qualitative assessment required was aided by searching the corpus for the given
nouns and then reading what they were about. This search-based aid was necessary
because many of the tokens are about technical terms, including
abbreviations. The other author was instructed to do the same. He was also given
the complete topic groupings instead of the shortened summaries presented in the
paper.

\section{Results}\label{sec: results}

\subsection{Explicit Security Considerations and Keywords}

The results are presented by going through the RQs consecutively. Thus,
regarding RQ.1, the two parsing rules caught only $779$ cases without explicit
security considerations despite the existence of the standardized section for
security considerations. Given that $779/n \times 100 \approx 9.2\%$, it can be
concluded that explicit security considerations are generally well-represented
in the RFCs sampled.

\begin{table}[th!b]
\centering
\caption{Correlations (Spearman) Between the Centrality Measures}
\label{tab: centrality correlations}
\begin{tabular}{llrrrr}
\toprule
&&\qquad\qquad 1. & 2. & 3. \\
\hline
1. & Weighted in-degree & 1.00 & 0.53 & 0.13 \\
2. & Betweenness & 0.53 & 1.00 & 0.21  \\
3. & Closeness & 0.13 & 0.21 & 1.00 & \\
\bottomrule
\end{tabular}
\end{table}

\begin{figure}[th!b]
\centering
\includegraphics[width=\linewidth, height=8cm]{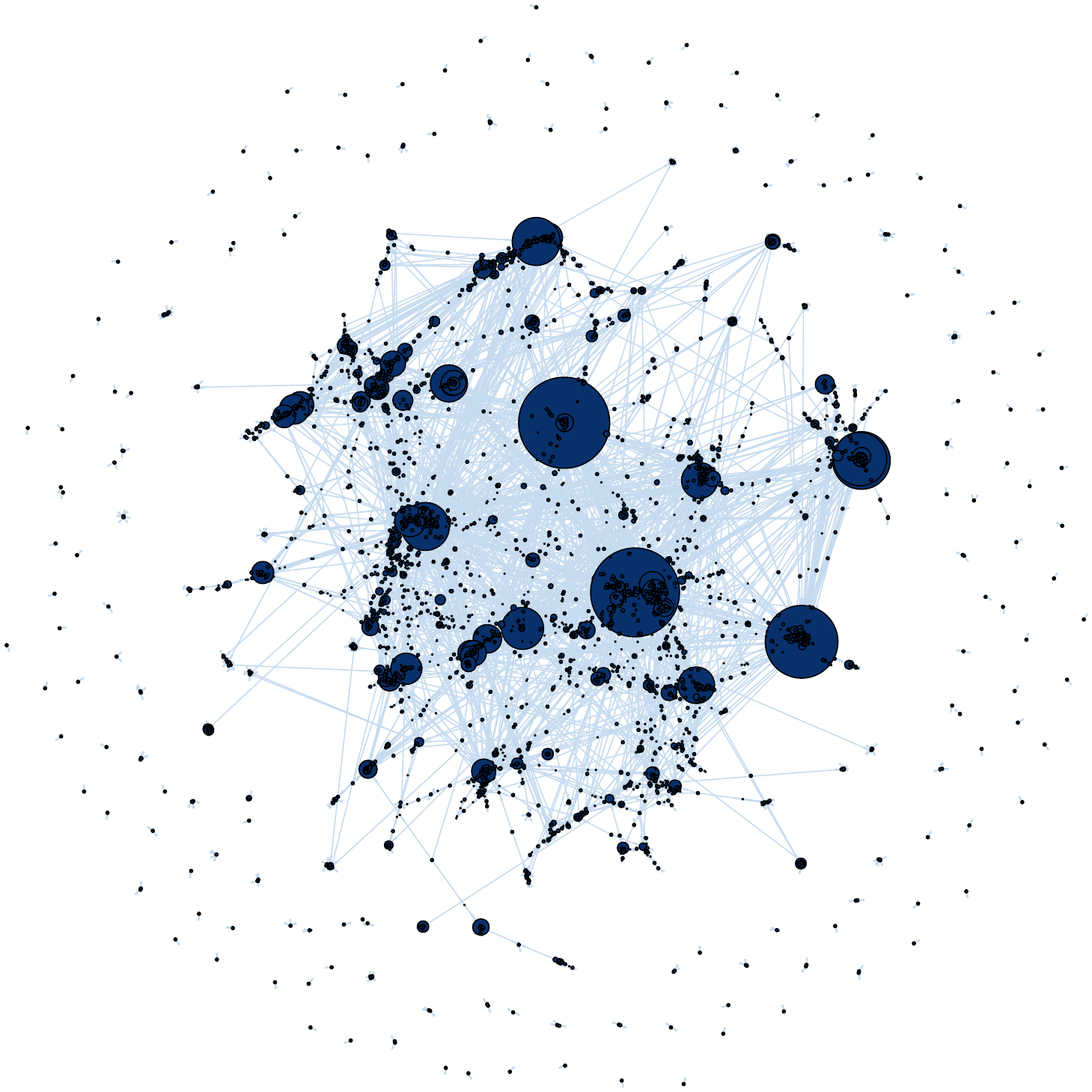}
\caption{The RFC-to-RFC Reference Network (sizes of vertices scaled by their weighted in-degrees)}
\label{fig: network}
\end{figure}

However, a similar conclusion does not apply with respect to the IETF's
standardized keywords. As was already noted in Subsection~\ref{subsec: parsing},
not a single upper-cased keyword appears in the security sections. With
lower-cased matching, which may contain also false positives, the order is: $18$
occurrences of ``should'', $14$ matches of ``shall'', $7$ cases of ``must'', $3$
occurrences of ``should not'', and $2$ matches of ``must not''. Given these
numbers and the guidelines~\cite{RFC8174}, it can be concluded that the security
considerations have seldom contained strict so-called
MoSCoW-style~\cite{Miranda11} requirement~specifications for security.

\begin{figure*}[th!b]
\centering
\includegraphics[width=14cm, height=10cm]{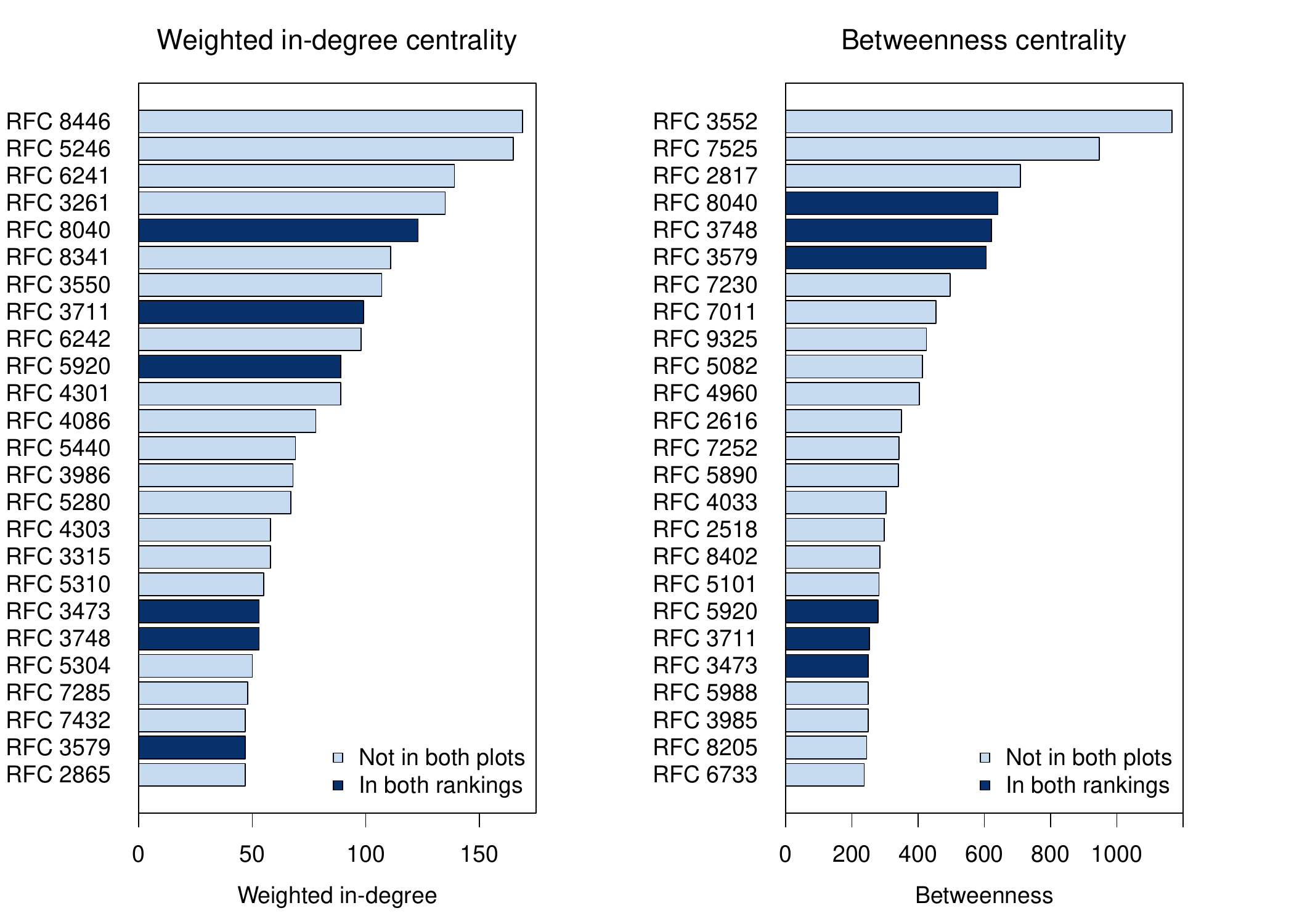}
\caption{Top-25 Referenced RFCs According to Two Centrality Measures}
\label{fig: centrality rankings}
\end{figure*}

\begin{figure*}[th!b]
\centering
\includegraphics[width=14cm, height=5cm]{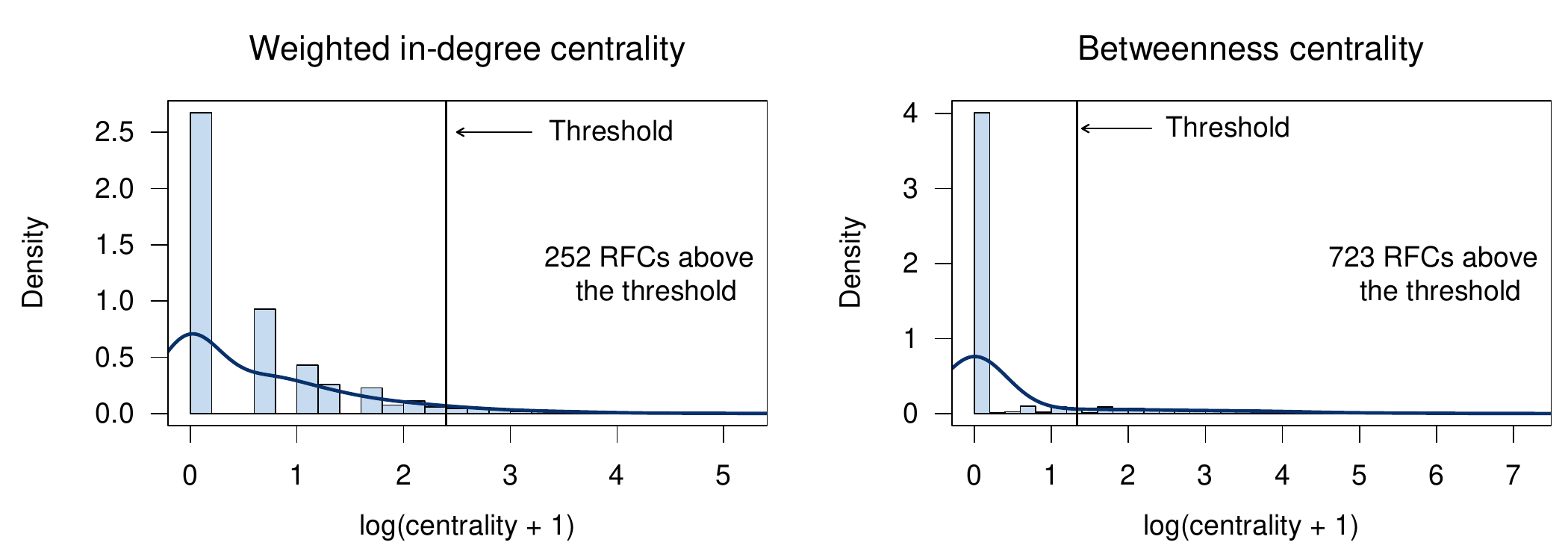}
\caption{Distributions of the Weighted In-degree and Betweenness Centrality Values}
\label{fig: centrality hist}
\end{figure*}

\begin{figure*}[th!b]
\centering
\includegraphics[width=14cm, height=10cm]{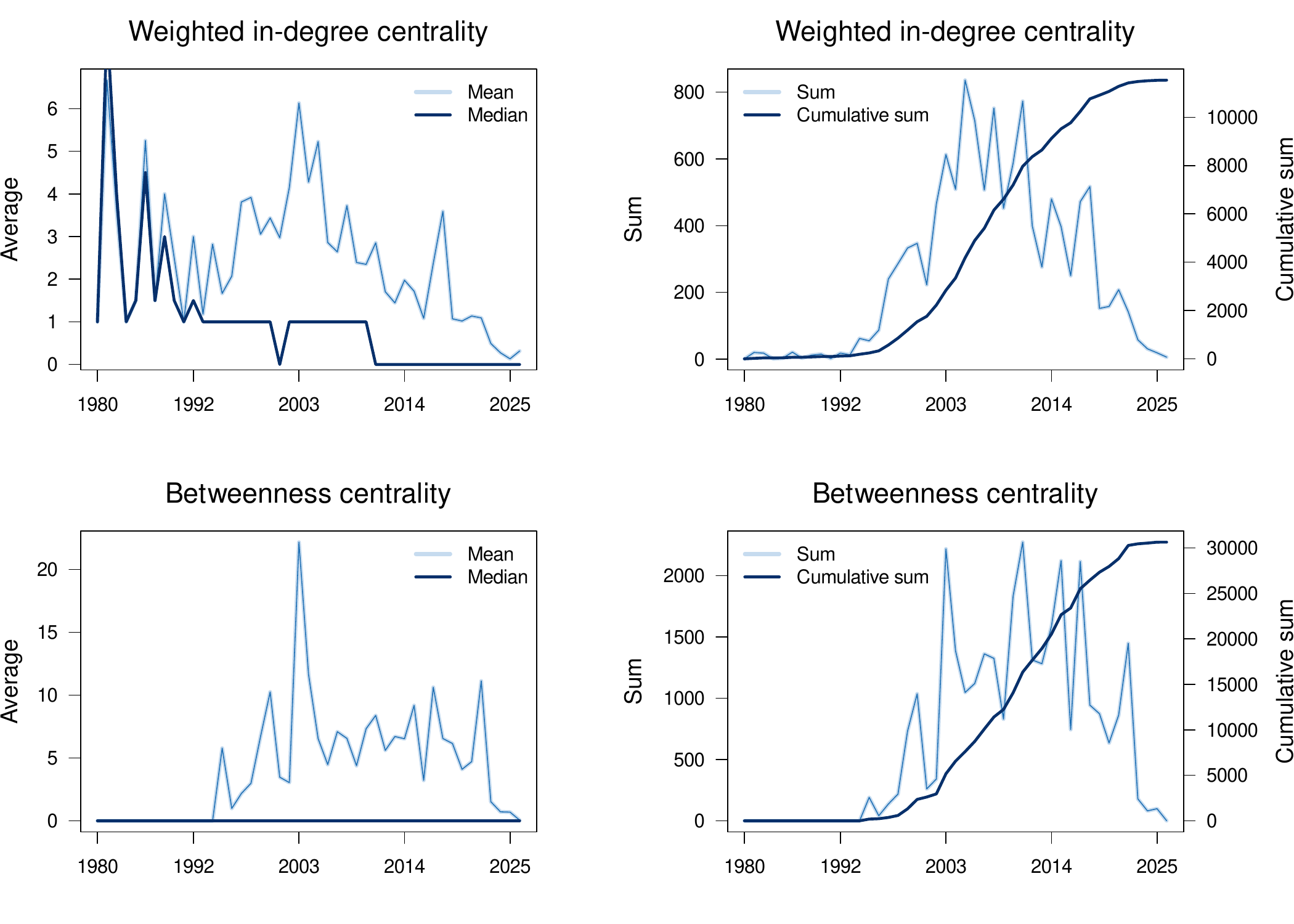}
\caption{Weighted In-degree and Betweenness Centrality Across Publication Years}
\label{fig: centrality years}
\end{figure*}

\subsection{Network Analysis Results}

Turning to RQ.2, it can be started by noting that the RFC-to-RFC referencing
network is sparse. In fact, its density---the ratio of edges to the theoretical
maximum number of edges---is only $0.0004$. In other words, many RFCs have
neither referenced other RFCs nor having been referenced by other RFCs. However,
as seen from the visualization in Fig.~\ref{fig: network}, a few of the RFCs
have been heavily referenced. Before continuing to elaborate these RFCs and the
visualization further, the correlations between the three centrality measures
should be assessed. These are shown in Table~\ref{tab: centrality
  correlations}. As can be seen, the weighted in-degrees only correlate fairly
with the betweenness centrality values. Thus, as per the discussion in
Subsection~\ref{subsec: network analysis}, the betweenness centrality is taken
alongside the weighted in-degrees for further examination.

Yet, to continue briefly with the weighted in-degrees, the two largest, almost
equally sized vertices in Fig.~\ref{fig: network} denote RFCs for the transport
layer security (TLS) protocol versions 1.2 and 1.3. Given the fundamental nature
of the TLS for the whole Internet's security, the two topmost places in terms of
the weighted in-degree ranking in Fig.~\ref{fig: centrality rankings} are hardly
surprising. The third and fourth places in the left-hand side plot of the figure
are worth also explicitly mentioning: these are specifications for the network
configuration protocol (NETCONF) and the session initiation protocol (SIP),
released in 2011 and 2002, respectively. The fifth place is taken by RFC
8040~\cite{rfc8040}, which is also about NETCONF. This RFC 8040 takes also the
fourth place in the betweenness centrality ranking shown in the right-hand side
plot of Fig.~\ref{fig: centrality rankings}. In general, however, there are only
six RFCs that appear in both top-25 centrality rankings. This observation
further supports the choice to include the betweenness centrality measure to
accompany the weighted in-degree values for constructing the sample.

The highest ranked RFC in terms of the betweenness centrality values is also
worth mentioning; it is RFC 3552, which is no more and no less than about
guidelines for writing the standardized sections for security
considerations~\cite{RFC3552}. This result aligns with the notion of betweenness
centrality as a measure on how frequently a vertex appears in the shortest paths
between other nodes. In other words, it could indeed be seen as something
central through which generic RFC security information passes.

However, the rankings are not particularly relevant for answering to RQ.2. In
this regard, Fig.~\ref{fig: centrality hist} displays the empirical probability
distributions of the two centrality measure values. Both are long-tailed but
both also indicate relatively interpretable cutoff points regarding their tails
and hence the most central RFCs. A total of $975$ RFCs exceed the thresholds
shown in the figure's two plots. Of these, $113$ are duplicates and thus the
heuristic answer to RQ.2 can be conveyed as a share $862/n \times 100 =
10.2\%$. Even with the subjectivity involved with scree plots, the answer seems
generally sensible also in terms of Fig.~\ref{fig: network}.

Regarding RQ.3, a further point to deduce from Fig.~\ref{fig: centrality
  rankings} is that many of the top-ranked RFCs are quite recent, relatively
speaking---keeping in mind the earlier remark about the first RFC being from
1969. The four visualizations in Fig.~\ref{fig: centrality years} shed further
light. By focusing on the weighted in-degrees for brevity, the two upper plots
indicate that the volume of references peaked during a period circa from
mid-1990s to mid-2010s. When focusing on the top-left plot in the figure, it can
also be concluded that the average reference amounts started to decline already
around 2005. Therefore, much of the growth trends in the top-right plot are
explained by the top-referenced RFCs in the earlier Fig.~\ref{fig: centrality
  rankings}. A further point is that the security-specific referencing trends
differ from those reported for all references, which tended to exhibit a slowly
increasing trend between 2001 and 2020 with about ten to fifteen RFC-specific
references~\cite[Fig.~7]{McQuistin21}. With these results and interpretations,
the topic modeling results can be presented next.

\subsection{Topic Modeling Results}\label{subsec: topic modeling results}

\begin{figure}[th!b]
\centering
\includegraphics[width=8cm, height=3cm]{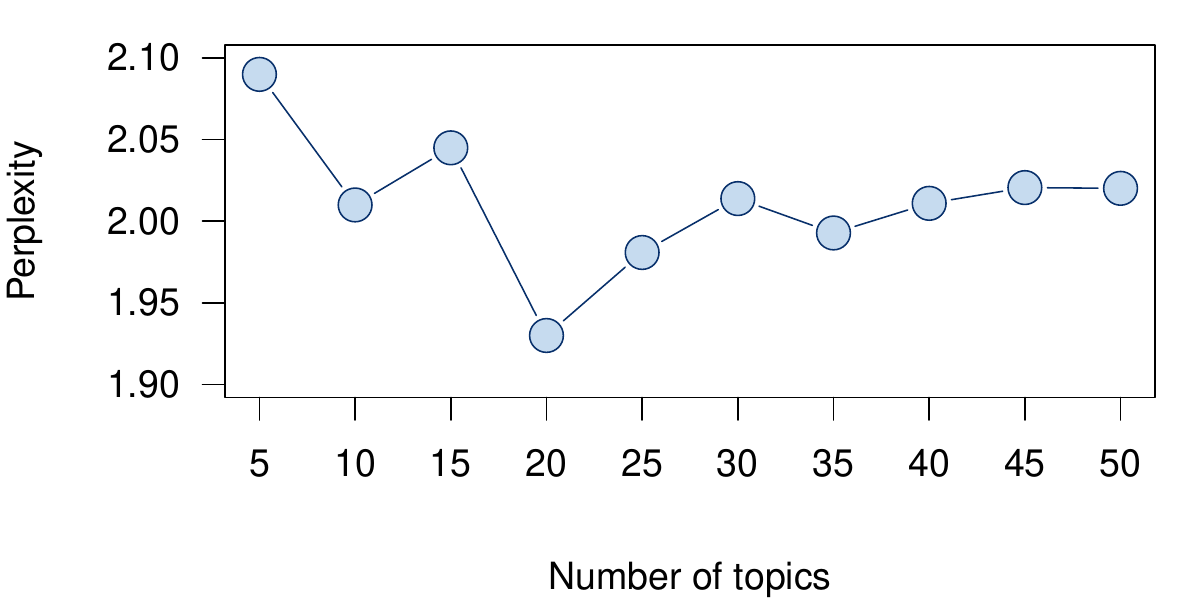}
\caption{Perplexity Values}
\label{fig: perplexity}
\end{figure}

\begin{figure*}[p!]
\centering
\includegraphics[width=\linewidth, height=20cm]{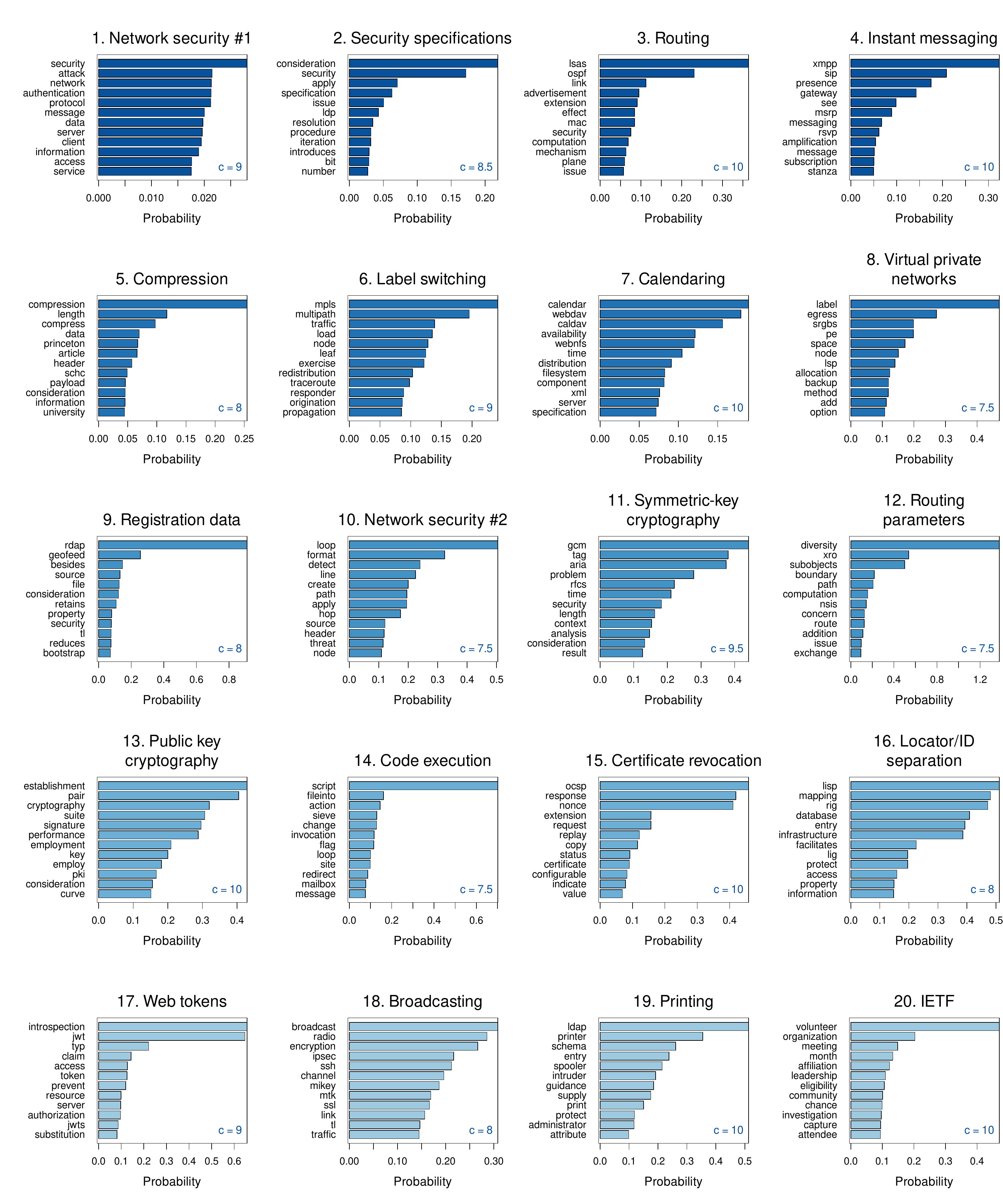}
\caption{The Top-12 Tokens in Each Twenty Topics}
\label{fig: topics}
\end{figure*}

The choice of $t = 20$ provides a good starting point for presenting the topic
modeling results. While, as said, interpretability was the primary criterion,
also the perplexity values shown in Fig.~\ref{fig: perplexity} support the
choice. Thus, the $20$ topics extracted and the top-12 nouns in each of them are
shown in Fig.~\ref{fig: topics}. The scalar $c$ at the bottom-right corner of
the figure's each plot refers to the mean of the interpretability and coherence
rankings done by the two authors independently and blinded from each other. As
can be observed, the values are generally high; interpretability and coherence
were generally perceived as being suitable. In fact, the minimum value is as
high as $7.5$. Furthermore, no disagreements were recorded between the two
authors.

\begin{figure}[th!b]
\centering
\includegraphics[width=\linewidth, height=10cm]{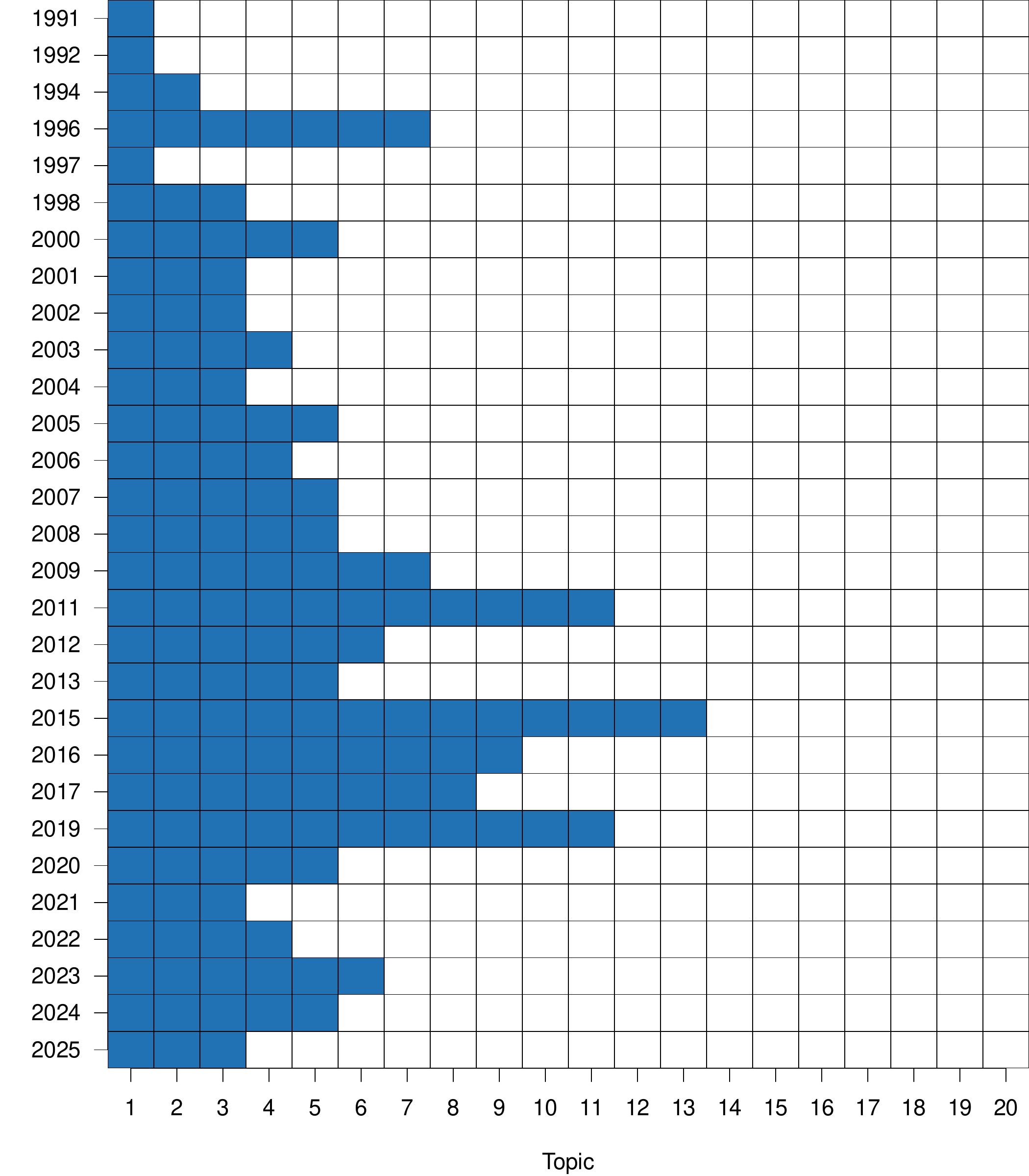}
\caption{Annual Presence of the Topics}
\label{fig: topics years}
\end{figure}

Before continuing, it should be noted that the top-12 nouns do not convey all
information that was used for the labeling of the topics. For instance, the
label for the eight topic cannot be easily deduced from the tokens shown in
Fig.~\ref{fig: topics}. Although not visible in the top-12 rankings, the topic
contains the tokens \textit{evpn} and \textit{mvpn}. Given that these are
lower-cased abbreviations for Ethernet VPN, where VPN stands for Virtual Private
Network, and Mobile VPN, the topic was labeled to represent VPNs.

Then, there are four generic topics: the topics 1 and 10 are about network
security general, the topic 2 is about terminology typical to the sections for
security considerations, and the topic 20 is about organizational matters
related to IETF and RFC drafting. As further seen from Fig.~\ref{fig: topics
  years}, the first topic is the only one occurring throughout the time period
observed. Also the second and third topics mostly occur each year. This
observation is understandable because the topics 1 and 2 are generic, whereas
the topic 3 is about routing, which is a core functionality of the Internet,
including with respect to security.

With the exception of the four generic topics and the two topics representing
cryptography, the topics do not seem to represent distinct security issues, such
as denial of service attacks, security issues with encapsulation and
fragmentation, spoofing, data injection, tampering, traffic analysis,
authentication and authorization, reconnaissance, eavesdropping, data injection,
replaying, firewall traversal, issues with network address translation, key
management insecurities, deprecated protocols, or something else. Rather---and
with respect to RQ.4, the topics seem to reflect distinct security issues in
distinct protocols. Excluding the first three topics already noted, also the
longitudinal evolution of the topics visualized in Fig.~\ref{fig: topics years}
can be interpreted to reflect the time periods when the protocols were actively
developed and standardized.

\section{Conclusion and Discussion}\label{sec: conclusion}

\subsection{Conclusion}

The paper examined security considerations explicitly discussed in RFCs
published by the IETF. Four research questions were examined. The answers to
these four RQs can be consecutively summarized as follows:
\begin{enumerate}
\item{The clear majority (over 90\%) of the RFCs observed have complied with RFC
  1543 by providing explicit sections for security considerations. However, only
  a very small minority have used the standardized keywords, such as MUST and
  SHOULD, originally specified in RFC 2119 from 1997. Indirectly, this
  observation indicates that security considerations have seldom been specified
  as explicit requirements. The lack of keywords used in the sections for
  security considerations is noteworthy also because the use of keywords in RFCs
  has generally grown since 2001~\cite{McQuistin21}. Their use has been
  recommended also for security considerations~\cite{White21}, but, according to
  the results, to no avail.}

\item{There are a few heavily referenced RFCs, although generally the RFC-to-RFC
  reference network is sparse. Particularly the protocols and associated RFCs
  for TLS stand out in terms of inward references they have gathered from other
  RFCs.}
\item{The volume of references peaked between circa mid-1990s to mid-2010s, and
  the heavily referenced RFCs are mostly quite recent. The result is interesting
  in a sense that the contrary could have been expected. In other words, the
  fundamental, the core RFCs published prior to mid-1990s could have been
  hypothesized to gather more references. The earlier point about TLS may
  explain also this longitudinal result.}
\item{The topics in the sections for security considerations are mostly
  protocol-specific instead of being about general, protocol-agnostic security
  topics, such as those related to denial of service attacks, authentication and
  authorization, or key management. With the exceptions of three general topics
  about network security in general, security specifications, and routing, the
  longitudinal evolution of the topics also tend to reflect the time periods
  during which the specific protocols were actively designed and standardized.}
\end{enumerate}

\subsection{Limitations}\label{subsec: limitations}

A notable yet still hypothetical limitation relates to potential parsing
errors. As was noted in Subsection~\ref{subsec: parsing}, complex within-RFC and
across-RFC referencing, including the use of appendices, is another notable
limitation. That said, there is a trade-off between these two limitations; by
trying to cover sections referenced, parsing errors would have likely
increased. As always with text mining, the topic modeling results can be
acknowledged to also present only a coarse picture of the actual security
considerations discussed in the RFCs. To address this limitation, qualitative
research can be recommended for further work.

\subsection{Closing Remarks}

With respect to RFCs and the IETF in general, the observation about
protocol-centered security topics is a positive finding in a sense that
repetition and duplication are seldom needed. After all, on one hand, there
already exists a RFC for a glossary of security terminology~\cite{RFC2828}. On
the other hand, it could also be argued that a survey RFC could be drafted for
summarizing the distinct protocol-specific security issues and presenting the
big picture. Another idea would be to contemplate whether the ``plugin-style''
standardization strategy used with some RFCs~\cite{Ruohonen26COMPUTER} might fit
also into the security context. Given what was said in the preceding
Subsection~\ref{subsec: limitations}, better and more robust meta-data could
also be hoped for. This point is familiar also from other tangentially related
contexts~\cite{Ruohonen26CLSR}.

Many security standards have been seen as ambiguous~\cite{Boyes24, Moyon21}. The
same criticism applies also to RFCs whose sometimes ambiguous requirements have
been seen as a source of security issues of various kinds~\cite{White21}. As was
discussed in the previous Subsection~\ref{sec: background}, it may be possible
to fix ambiguity issues later on, but even then, the fixes, patches, and
extensions should be unambiguous even in case the original RFCs were
not. Further qualitative research is required for deducing whether and how
ambiguous the security considerations expressed really are. Such line of
research would align well with another topic worth investigating, the IETF's
standardization philosophy.

The 1990s architectural principles for the Internet, as specified in RFC
1958~\cite{RFC1958}, emphasized the so-called ``principle of good enough'' that
has been adopted also for software engineering~\cite{Bach03}. For instance, it
was written that in ``many cases it is better to adopt an almost complete
solution now, rather than to wait until a perfect solution can be
found''~\cite{RFC1958}.  The security ``mechanism described herein is not
perfect and does not need to be perfect'', as one RFC noted~\cite{RFC7166}. With
these quotations in mind, further research is needed about the fundamental
question; about whether the sections for security considerations have actually
helped at improving security of the~Internet.

\balance
\bibliographystyle{apalike}

\end{document}